\documentclass[journal=jacsat,manuscript=article]{achemso}

\usepackage{amsmath}
\usepackage{epstopdf}
\usepackage{graphicx}
\usepackage{dcolumn}
\usepackage{bm}
\usepackage{braket}
\usepackage{gensymb}
\usepackage{array}
\usepackage[utf8]{inputenc}
\usepackage[T1]{fontenc}
\usepackage{mathptmx}

\newcommand{\bs}{\boldsymbol}

\title{Ultrafast Fluorescence Depolarization in Conjugated Polymers}

\author{Isabel Gonzalvez Perez}
\email{isabel.gonzalvez@chem.ox.ac.uk}
\affiliation{Department of Chemistry, Physical and Theoretical Chemistry Laboratory, University of Oxford, Oxford, OX1 3QZ, United Kingdom}

\author{William Barford}
\email{william.barford@chem.ox.ac.uk}
\affiliation{Department of Chemistry, Physical and Theoretical Chemistry Laboratory, University of Oxford, Oxford, OX1 3QZ, United Kingdom}

\begin{document}

\begin{abstract}
We report on large-scale simulations of intrachain exciton dynamics in poly(para-phenylene vinylene). Our coarse-grained model describes Frenkel exciton coupling to both fast, quantized C-C bond vibrations and slow, classical torsional modes. We also incorporate  system-bath interactions.
The dynamics are simulated using  the Time Evolution Block Decimation method, which avoids the failures of the Ehrenfest approximation to describe decoherence processes and  nonadiabatic interstate conversion. System-bath interactions are modeled using quantum trajectories and Lindblad quantum jump operators.
We find that following photoexcitation, the quantum mechanical entanglement of the exciton and C-C bond phonons causes exciton-site decoherence. Next, system-bath interactions cause the stochastic collapse of high-energy delocalized excitons onto chromophores. Finally, torsional relaxation causes additional exciton-density localization.
We relate these dynamical  processes to the predicted fluorescence depolarization and extract the timescales corresponding to them.
\end{abstract}

\vfill\pagebreak

A wide range of time-resolved spectroscopic techniques, including fluorescence depolarization\cite{Grage03,Ruseckas05,Wells07,Dykstra09}, three-pulse photon-echo\cite{Dykstra05,Yang05, Wells08,Sperling08} and coherent electronic two-dimensional spectroscopy\cite{Consani15}, indicate that exciton dynamics in conjugated polymers spans multiple and overlapping timescales.
Some of the timescales extracted from these experiments, ranging from a few femtoseconds to sub-ns, are listed in Table \ref{ta:2}.
Dynamics on timescales exceeding ca.\ 20 ps are usually assumed to be associated with interchromophore exciton diffusion in the condensed phase, while dynamics on timescales of less than ca.\ 10 fs are so fast, they must be occur from intrinsic intrachain processes.
Dynamics on intermediate timescales, however,  arise from  coupling to torsional modes and  system-bath interactions.
Wells and Blank's\cite{Wells08} analysis of two-color three-pulse echo peak shifts concluded that for P3HT in chloroform solution the driving mechanism for exciton relaxation within 200 fs is a coupling to overdamped torsional modes. On the other hand, for MEH-PPV Yang \textit{et al.}\cite{Dykstra05} had previously shown that the ultrafast dynamics in less than 30 fs corresponds to a stochastic process of localization.
Timescales of tens of ps have been ascribed to a collective motion of an exciton and the rotational degrees of freedom\cite{Albu13}, and indeed can be explained by intrachain exciton diffusion driven by stochastic torsional fluctuations\cite{Tozer15, Binder20c}.

\begin{table}
\small\centering
{\renewcommand{\arraystretch}{1.2}
\begin{tabular}{|p{2.5cm}|p{2.5cm}|p{8cm}|p{1.5cm}|}
\hline
Polymer & State &  Timescales & Citation \\
\hline
MEH-PPV & Solution & $\tau_1 = 50 $ fs, $\tau_2 = 1-2 $ ps & Ref\cite{Ruseckas05} \\
\hline
MEH-PPV & Solution & $\tau_1 = 5-10 $ fs, $\tau_2 = 100-200 $ fs  & Ref\cite{Collini09} \\
\hline
PDOPT & Film & $\tau = 0.5 - 4 $ ps & Ref\cite{Westenhoff06} \\
\hline
PDOPT & Solution & $\tau_1 < 1 $ ps, $\tau_2 = 15 -23 $ ps & Ref\cite{Westenhoff06} \\
\hline
P3HT & Film & $\tau_1 = 300 $ fs, $\tau_2 = 2.5 $ ps, $\tau_3 = 40 $ ps & Ref\cite{Westenhoff06} \\
\hline
P3HT & Solution & $\tau_1 = 700 $ fs, $\tau_2 = 6 $ ps, $\tau_3 = 41 $ ps, $\tau_4 = 530 $ ps & Ref\cite{Banerji11} \\
\hline
P3HT & Solution & $\tau_1 = 60 - 200 $ fs, $\tau_2 = 1-2 $ ps, $\tau_3 = 14-20 $ ps & Ref\cite{Busby11} \\
\hline
P3HT & Solution & $\tau_1 < 100$ fs, $\tau_2 = 1 - 10$  ps & Ref\cite{Wells07} \\
\hline
\end{tabular}}
\caption{Some of the excitonic dynamical timescales observed in conjugated polymers.}
\label{ta:2}
\end{table}

In this letter, we report on large-scale simulations of intrachain exciton dynamics in poly(para-phenylene vinylene) that can explain the sub-ps  processes. Our model describes exciton coupling to both fast, quantized C-C bond vibrations and slow, classical torsional modes. We also incorporate system-bath interactions. We relate these dynamical  processes to the predicted fluorescence depolarization and extract the timescales corresponding to them. Crucially, because the exciton dynamics spans 10 to 1000 femtoseconds  and over 100 monomers, coarse-grained models of excitons and nuclei must be employed. Furthermore, since the first two processes to be discussed are associated with  exciton decoherence and nonadiabatic energy relaxation via multiple potential energy surfaces, an Ehrenfest treatment of the nuclear dynamics is not appropriate.

The Ehrenfest method fails to describe ultrafast dynamics because of its two key  assumptions\cite{Horsfield06,Nelson20}. The first  is to treat the nuclei classically. This means that nuclear quantum tunneling and zero-point energies are neglected, and that exciton-polarons are not correctly described. The second assumption is that the total wavefunction is a product of the electronic and nuclear wavefunctions. This means that there is no entanglement between the  electrons and nuclei,  and so the nuclei cannot cause decoherence of the electronic degrees of freedom. A simple product wavefunction also implies that the nuclei  move in a mean field potential determined by the electrons. This means that a splitting of the nuclear wave packet when passing through a conical intersection or an avoided crossing does not occur, and that  there is an incorrect description of energy transfer between the electronic and nuclear degrees of freedom.

To avoid the failures of the Ehrenfest approximation in describing ultrafast process we use the Time Evolution Block Decimation (TEBD) method to numerically evolve the system's state vector under the action of the Frenkel-Holstein model in which  the fast C-C bond vibrations are fully quantized. (The TEBD method is briefly  described in the SI and implemented in the  Tensor Network Theory Library\cite{Al-assam17}.)

The fate of the exciton is controlled by a series of fascinating and complex quantum mechanical processes. Before describing this, however, we need to establish the optically prepared state. For a conformationally disordered polymer, the delocalization of the exciton center-of-mass wavefunction is determined by the disordered Frenkel model (as given by Eq.\ (1) in the SI). As described elsewhere\cite{Book,Barford14b} and in the SI, exciton eigenstates may be classified as low-energy local exciton ground states (LEGSs) and high-energy quasiextended exciton states (QEESs). LEGSs are non-overlapping, space-filling and quasinodeless. Their spatial extent defines chromophores. In contrast, as illustrated in Fig.\ 2 of the SI, QEESs are spatially extended, nodeful and spatially overlap lower-lying LEGSs.  In general, depending on the laser pulse, an initial state will be a non-stationary state composed of a linear superposition of these Frenkel model eigenstates.

The initial dynamical process following photoexcitation is the quantum mechanical correlation of the exciton with phonons associated with high-frequency C-C bond vibrations. Since the C-C bond vibrations are fast (one period is 21 fs) they must be treated  quantum mechanically, which we do using the fully quantized Frenkel-Holstein model (as given by Eq.\ (7) in the SI).
As shown in Fig.\ 7 of ref\cite{Mannouch18},
the instantaneous force on the nuclei  causes local nuclear displacements and the creation of an exciton-polaron within 10 fs of photoexcitation.

Crucially, the system wavefunction, which at photoexcitation is a direct product of the exciton and nuclear wavefunctions, is now a quantum mechanically entangled exciton-nuclear wavefunction. As a consequence, the exciton decoheres\cite{Schlosshauer07}, meaning that there is a loss of off-diagonal long range order (ODLRO) in the exciton's reduced density matrix. To quantify this statement, we introduce the coherence correlation function\cite{Kuhn97,Smyth12}
\begin{equation}\label{Eq:17a}
C_n^{\textrm{coh}}(t) = \sum_m \left| \rho_{m,m+n} \right|,
\end{equation}
where $\rho_{m,m'}$ is the exciton reduced density matrix obtained by tracing over the vibrational degrees of freedom.
Here, the density matrix is defined within the local Frenkel exciton basis $\{|m\rangle \}$, where the index $m$ labels a moiety (equivalent to a site in the Frenkel model).
The loss of site-basis coherence is further quantified by the coherence length, defined by
\begin{equation}\label{Eq:18}
  L_{\textrm{coh}}  = \sum_n C_n^{\textrm{coh}}.
\end{equation}
As illustrated in Fig.\ 1, $L_{\textrm{coh}}$ decays to ca.\ 4 moieties in ca.\ 10 fs, reflecting the localization of exciton coherence resulting from the short range exciton-phonon correlations.
\begin{figure}\label{Fi:1}
\includegraphics[width=0.7\textwidth]{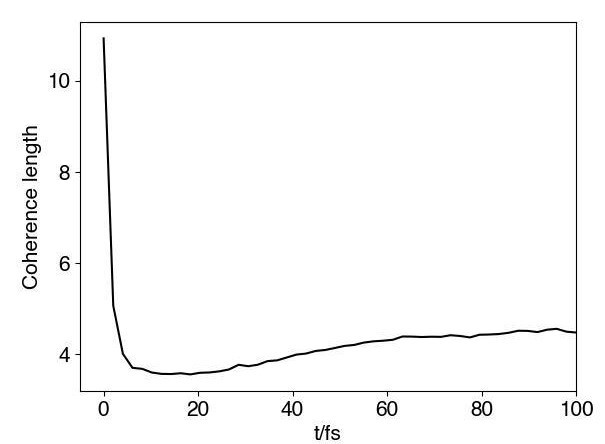}
\caption{The exciton coherence length $ L_{\textrm{coh}}$ (in number of moieties) as a function of time in PPV chains of 40 moieties. These results are calculated from an ensemble of delocalized high-energy states (i.e., QEESs) at $t=0$.
}
\end{figure}

The second dynamical process is associated with system-bath interactions. For an exciton to dissipate energy it must  first couple to fast internal degrees of freedom (as just described) and then these  degrees of freedom must couple to the environment to expell heat. For a low-energy exciton (i.e., a LEGS) this process will cause adiabatic relaxation on a single potential energy surface, forming a vibrationally relaxed state\cite{Tretiak02,Bittner03,Sterpone08,Leener09}.
However, for a kinetically hot exciton (i.e., a QEES)  this relaxation is through a dense manifold of states and is necessarily a nonadiabatic interstate conversion between different potential energy surfaces. As already mentioned, the Ehrenfest approximation fails to correctly describe this process\cite{Tozer12}, so again it is necessary to treat the exciton and fast C-C bond vibrations on the same quantum mechanical footing.

Dissipation of energy from an open quantum system arising from system-environment coupling is commonly described by a Lindblad master equation\cite{Breuer02,Schlosshauer07}, as discussed in the SI.
In practice, a direct solution of the Lindblad master equation is usually prohibitively expensive, as the size of Liouville space scales as the square of the size of the associated Hilbert space.
Instead, Hilbert space scaling can be maintained by performing ensemble averages over quantum trajectories (evaluated via the TEBD method), where the action of the Linblad dissipator is modeled by quantum jumps.\cite{Daley14}

Ultrafast exciton-site decoherence  occurs via the coupling of the exciton to fast internal degrees of freedom, namely the C-C bond vibrations. We note that this coupling does not cause exciton density localization. However, dissipation of energy to the environment causes an exciton in a higher energy QEES to relax onto a lower energy LEGS (i.e., onto a chromophore) and thus the exciton density becomes localized.
The spatial extent of the exciton density, averaged over an ensemble of quantum trajectories and different realizations of the disorder, is quantified by the localization length, defined by\cite{Book}
\begin{equation}\label{Eq:17}
L_{\text{loc}}= 5.6\left( \langle L^2 \rangle - \langle L \rangle^2\right)^{1/2}.
\end{equation}
Fig.\ 2 shows the time dependence of  $L_{\text{loc}}$ with an external dissipation time $ \gamma^{-1} = 100$ fs.
\begin{figure}\label{Fi:2}
\includegraphics[width=0.7\textwidth]{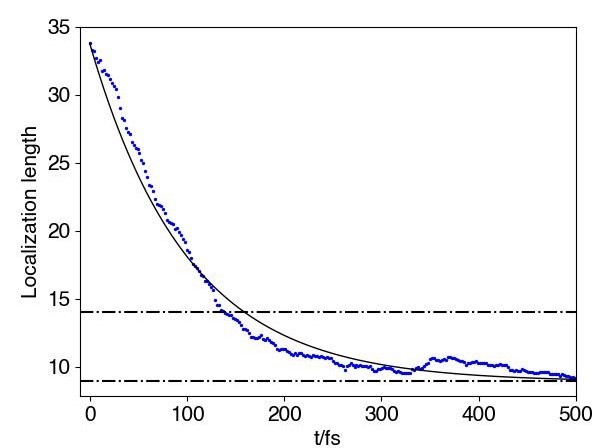}
\caption{The exciton localization length $ L_{\textrm{loc}}$ (in number of moieties) as a function of time in PPV chains of 40 moieties. These results are calculated from an ensemble of delocalized high-energy states (i.e., QEESs) at $t=0$.
The horizontal dashed line at 14 moieties  corresponds to the average chromophore size in the absence of exciton-torsional coupling (i.e., to $L_{\text{loc}}$ averaged over all LEGSs). The asymptotic value of  9 moieties is a consequence of additional localization caused by torsional relaxation.
The curve is a one-exponential fit with a time constant of 100 fs.
}
\end{figure}

Figure 2 is obtained by averaging over an ensemble of quantum trajectories starting from a distribution of different QEESs. To understand the physical process of localization onto a chromophore, Fig.\ 12 of ref\cite{Mannouch18} illustrates the exciton density of a single quantum trajectory for a photoexcited QEES. At a time  ca.\ 20~fs  a `quantum jump' caused by the stochastic application of a Lindblad jump operator causes the exciton to  localize onto a LEGS, i.e., the high-energy extended state has randomly localized onto a chromophore because of a `measurement' by the environment.

By dissipating energy into the environment on sub-ps timescales, hot excitons relax into localized LEGSs, i.e., onto chromophores.
In our model the final intrachain relaxation and localization process now takes place, namely exciton-polaron formation via  coupling to the torsional degrees of freedom. For this relaxation to occur bond rotations must be allowed, which means that this process is highly dependent on the precise chemical structure of the polymer and its environment.
Assuming that bond rotations are not sterically hindered, their coupling to the excitons is conveniently modeled by supplementing the Frenkel-Holstein model by $\hat{H}_{\textrm{rot}}$\cite{Barford18} (as given by Eq.\ (9) of the SI).

Unlike C-C bond vibrations, being over 10 times slower torsional oscillations can be treated classically\cite{Barford18}. Furthermore, since we are now concerned with adiabatic relaxation on a single potential energy surface, we may employ the Ehrenfest dynamics, meaning that the exciton dynamics is described by the time-dependent Schr\"odinger equation while the torsional modes are subject to Newton's equation of motion with a mean field given by the exciton density (as described further in the SI).

An exciton delocalized along a polymer chain in a chromophore couples to multiple rotational oscillators resulting in collective oscillator dynamics.
Since the torsional motion is slow, the self-trapped exciton-polaron  is `heavy' and in the under-damped regime becomes self-localized on a timescale of a single torsional period (i.e., ca.\ $200$ fs in our calculations). In this limit the relaxed staggered  angular displacement mirrors the exciton density. Thus, the exciton is localized precisely as for a `classical' Landau polaron and is spread over ca.\ 9 moieties\cite{Barford18}, as shown by the asymptotic behavior of $L^{\text{loc}}$ in Fig.\ 2.

The horizontal dashed line at 14 moieties in Fig.\ 2 corresponds to the average chromophore size in the absence of torsional relaxation, i.e., to $L_{\text{loc}}$ averaged over all LEGSs. The  asymptotic value of ca.\ 9 moieties is a consequence of the additional density localization caused by torsional planarization. (Some authors argue that planarization causes exciton delocalization\cite{Westenhoff06}. We discuss this point in the SI.)

For general polymer conformations, the loss of ODLRO (or the localization of the exciton coherence function) causes a reduction and  rotation of the transition dipole moment.
The rotation is quantified by the fluorescence anisotropy, defined by\cite{Lakowicz06}
\begin{equation}
\label{eq:r}
r = \frac{I_{\parallel}-I_{\perp}}{I_{\parallel}+2I_{\perp}},
\end{equation}
where $I_{\parallel}$ and $I_{\perp}$ are the intensities of the fluorescence radiation polarised parallel and perpendicular to the incident radiation, respectively.
For an arbitrary state of a quantum system, $|\Psi\rangle$, the integrated fluorescence intensity polarised along the $x$-axis is related to the $x$-component of the transition dipole operator, $\hat{\mu}_{x}$, by
\begin{eqnarray}\label{Eq:20}
\label{t}
I_{x} \propto  \sum_{v}\left|\langle\Psi|\hat{\mu}_{x}|\textrm{GS},v\rangle\right|^{2},
\end{eqnarray}
where $|\textrm{GS},v\rangle$ corresponds to the system in the ground electronic state, with the nuclear degrees of freedom in the state characterised by the quantum number $v$.

The averaged fluorescence anisotropy is defined by
\begin{equation}
\label{eq:intens_av}
\langle r\left(t\right)\rangle = 0.4\times\frac{\sum_{i}I_{i}\left(t\right)r_{i}\left(t\right)}{\sum_{i}I_{i}\left(t\right)},
\end{equation}
where $I_{i}\left(t\right)$ is the total fluorescence intensity and $r_{i}\left(t\right)$ is the fluorescence anisotropy, associated with a particular conformation $i$ at time $t$. The factor of 0.4 is included on the assumption that the polymers are oriented uniformly in the bulk material.\cite{Lakowicz06} Fig.\ 3 shows the simulated $\langle r\left(t\right)\rangle$ for  an ensemble of initial high energy QEESs  over an ensemble of conformationally disordered polymers.

\begin{figure}\label{Fi:3}
\includegraphics[width=0.7\textwidth]{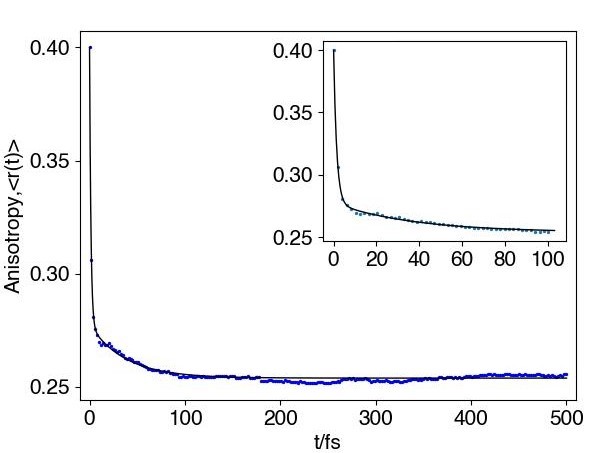}
\caption{The fluorescence anisotropy $\langle r\left(t\right)\rangle$ as a function of time calculated from an ensemble of delocalized high-energy states (QEESs) at $t=0$.
The curve is a two-exponential fit given by Eq.\ (\ref{Eq:10}).}
\end{figure}

To understand the sub-10 fs depolarization, it is instructive to express Eq.\ (\ref{Eq:20}) as
\begin{eqnarray}\label{Eq:21}
I_{x} \propto  \sum_{m,n} s_m^x s_n^x \rho_{mn},
\end{eqnarray}
where $s^x_m$ is the $x$-component of the unit vector for the $m$th moiety and $\rho_{mn}$ is the exciton reduced density matrix.
Then, using Eq.\ (\ref{Eq:17a}), Eq.\ (\ref{Eq:18}), and Eq.\ (\ref{Eq:21}),  we observe  that the emission intensity, $I_{x}$, is related to the coherence length, $L_{\textrm{coh}}$. Thus, not surprisingly, the initial dynamics of $\langle r\left(t\right)\rangle$ resembles that of $L_{\textrm{coh}}(t)$ shown in Fig.\ 1. In particular, we observe a loss of fluorescence anisotropy within 10 fs, mirroring the reduction of $L_{\textrm{coh}}$ in the same time.
This predicted loss of fluorescence anisotropy within 10 fs has been observed experimentally, as shown in Fig.\ 7 of ref\cite{Wells07}.

Slower sub-ps depolarization
occurs because of subsequent exciton density localization as discussed above, first by stochastic localization and second by coupling to torsional modes.
This density localization also causes a change in the transition dipole moments\cite{Ruseckas05}.

The inset of Fig.\ 3 indicates that the majority of sub-ps fluorescence depolarization occurs within a few fs. This is confirmed by our best functional fit of the data, which is of the form
\begin{equation}
\label{Eq:10}
\langle r\left(t\right)\rangle = r_1\exp(-t/\tau_1) + r_2\exp(-t/\tau_2)  + r_{\infty},
\end{equation}
where $\tau_1= 2$ fs and $\tau_2= 37$ fs; and $r_1 = 0.122$, $r_2 = 0.024$  and $r_{\infty}=0.254$.
The slower, less significant depolarization, has a time constant of 37 fs.

In summary, we have simulated the various sub-ps exciton dynamical processes in conjugated polymers. First, following photoexcitation, the initial dynamical process  is the quantum mechanical correlation of the exciton and phonons associated with high-frequency C-C bond vibrations. This quantum  entanglement causes exciton-site decoherence, which   manifests itself as sub-10 fs fluorescence depolarization. Next, the energy that is transferred from the exciton to the nuclei is dissipated into the environment on a timescale determined by the strength of the system-bath interactions. For a hot exciton (i.e., a QEES) the system-bath interactions cause the entangled exciton-nuclear wavefunction to stochastically `collapse' into a particular LEGS, causing the exciton density to be localized on a `chromophore'. Finally, the fate of an exciton on a chromophore is now strongly dependent on the polymer chemical structure and the type of environment. For underdamped, freely rotating moieties, the coupling of the exciton to the low-frequency classical torsional modes creates an exciton-polaron, with associated  planarization and  exciton-density localization.

In our simulations, the first of these  processes is indentifiable by a rapid fluorescence depolarization with a time constant of ca.\ 2 fs. However, because the last two processes both correspond to intrachain exciton density localization and occur on similar timescales, we have only been able to extract one time constant for the associated depolarization, namely 37 fs. Experimentally, it is observed that there is significant further depolarization on post-ps timescales; the most significant component being caused by interchromophore exciton diffusion.

\begin{acknowledgement}
I.G.P. acknowledges the EPSRC Centre for Doctoral Training, Theory and Modelling in Chemical Sciences under Grant No. EP/L015722/1 for financial support.
\end{acknowledgement}

\vfill\pagebreak

\begin{suppinfo}

Table of contents:
\begin{enumerate}
\item{Excitons in Conjugated Polymers}
\item{Fast C-C Bond Vibrations}
\item{Slow Bond Rotations}
\item{Open Quantum Systems}
\item{Numerical Techniques}
\end{enumerate}

\section{1.\ Excitons in Conjugated Polymers}\label{se:3}

\subsection{The Frenkel model}\label{se:3.1}

An exciton is a Coulombically bound electron-hole pair formed by the linear combination of electron-hole excitations
(for further details see\cite{Book, Abe93, Barford13}). In a one-dimensional conjugated polymer an exciton is described by the two-particle wavefunction,
  $\Phi_{mj}(r,R) = \psi_m(r)\Psi_j(R)$.
The relative wavefunction, $\psi_m(r)$,  describes a particle bound to a screened Coulomb potential, where $r$ is the electron-hole separation and $m$ is the principal quantum number.

Frenkel excitons, for which $m=1$, are the primary photoexcited states of light-emitting conjugated polymers. The center-of-mass wavefunction, $\Psi_j(R)$, is described by the Frenkel Hamiltonian,
\begin{eqnarray}\label{Eq:1}
  \hat{H}_{F} &=& \sum_{n=1}^{N}\epsilon_n  \hat{N}_{n} +\sum_{n=1}^{N-1}J_n \left( \ket{n}\bra{n+1} +  \ket{n+1}\bra{n}\right),
\end{eqnarray}
where $N$ is the number of moieties, $n = (R/d)$  labels a moiety, and $d$ is the intermoiety separation.
The energy to excite a Frenkel exciton on moiety $n$ is $\epsilon_{n}$, where
 $\hat{N}_{n} = \ket{n}\bra{n}$
is the Frenkel exciton number operator.

In principle,  excitons delocalize along the chain via two mechanisms\cite{Barford09a,Barford13}. First, for even-parity (odd $m$) singlet excitons  there is a Coulomb-induced (or through space) mechanism. This is the familiar mechanism of F\"orster  energy transfer.
In the point-dipole approximation
\begin{equation}\label{Eq:4}
    J_{DA}^{mn} = \frac{\kappa_{mn} \mu_0^2}{4\pi\varepsilon_{r}\varepsilon_{0} R_{mn}^3},
\end{equation}
where  $\mu_0$ is the transition dipole moment of a single moiety and ${R}_{mn}$ is the distance between the moieties $m$ and $n$.
$\kappa_{mn}$ is the orientational factor,
\begin{equation}\label{Eq:40}
\kappa_{mn} = \bs{\hat{r}_m}\cdot\bs{\hat{r}_{n}} - 3(\bs{\hat{R}_{mn}}\cdot\bs{\hat{r}_{m}})(\bs{\hat{R}_{mn}}\cdot\bs{\hat{r}_{n}}),
\end{equation}
where $\bs{\hat{r}_{m}}$ is a unit vector parallel to the dipole on moiety $m$ and $\bs{\hat{R}_{mn}}$ is a unit vector parallel to the vector joining moieties $m$ and $n$. For colinear moieties, the nearest neighbor through space transfer integral is
\begin{equation}
    J_{DA} = -\frac{2 \mu_0^2}{4\pi\varepsilon_{r}\varepsilon_{0} d^3}.
\end{equation}

\begin{figure}[t]
\includegraphics[width=0.7\textwidth]{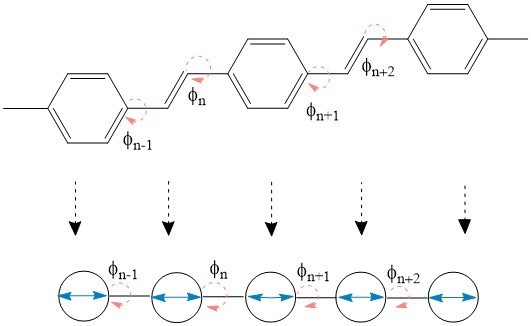}
\caption{A graphical depiction of our coarse-grained model of PPV.
Each site corresponds to a moiety along the polymer chain, with the connection between sites characterised by the torsional (or dihedral) angle, $\theta_n = (\phi_{n+1}-\phi_n)$. The nuclear degrees of freedom, both vibrational (blue) and torsional (red), are represented via  harmonic oscillators at each site, which are in turn locally coupled to the exciton.}
\label{Fi:1}
\end{figure}

Second, for all excitons there is a super-exchange (or through bond) mechanism, whose origin lies in a virtual fluctuation from a Frenkel exciton on a single moiety to a charge-transfer exciton spanning two moieties back to a Frenkel exciton on a neighboring moiety. The energy scale for this process, obtained from second order perturbation theory\cite{Book}, is
\begin{equation}\label{Eq:5}
    J_{SE}(\theta) \propto -\frac{t(\theta)^2}{\Delta E},
\end{equation}
where $ t(\theta)$  is proportional to the overlap of  $p$-orbitals neighboring a bridging bond, i.e., $t(\theta) \propto \cos \theta$ and  $\theta$ is the torsional (or dihedral) angle between neighboring moieties. $\Delta E$ is the difference in energy between a charge-transfer and Frenkel exciton.
The total exciton transfer integral is  thus
\begin{equation}\label{Eq:4}
J_{n} = J_{\text{DA}} + J_{\text{SE}}^0\cos^{2}{\theta_{n}}.
\end{equation}

Eq.\ (\ref{Eq:1}) represents a `coarse-graining' of the exciton degrees of freedom. The key assumption is that we can replace the atomistic detail of each  moiety and replace it by a `coarse-grained' site, as illustrated in Fig.\ 4.

\subsection{Local exciton ground states (LEGSs) and quasiextended exciton states (QEESs) }\label{se:3.2}

\begin{figure}[h]
\centering
\includegraphics[width=0.6\textwidth]{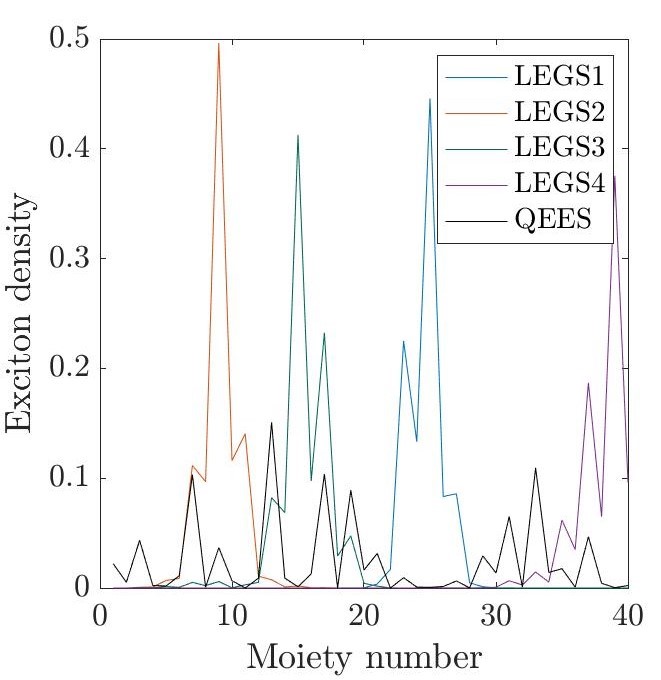}
  \caption{The density of the four local exciton ground states (LEGSs) and one of the quasiextended exciton states (QEES) for one particular static conformation of a PPV polymer chain of 40 moieties (i.e., 20 monomers).
}\label{Fi:2}
\end{figure}

Polymers are  subject to conformational, chemical and environmental disorder. Disorder localizes a quantum particle and determines their energetic and spatial distributions.  Malyshev and Malyshev\cite{Malyshev01a,Malyshev01b} further observed that in one-dimensional systems there are a class of states in the low energy tail of the density of states that are superlocalized,
named local exciton ground states (LEGSs\cite{Malyshev01a,Malyshev01b,Makhov10}). LEGSs are essentially nodeless, nonoverlapping wavefunctions that together spatially span the entire chain. They are \emph{local} ground states, because for the individual parts of the chain that they span there are no lower energy states. A consequence of the essentially nodeless quality of LEGSs is that the square of their transition dipole moment scales as their size\cite{Makhov10}. Thus, LEGSs define  chromophores (or spectroscopic segments), namely the irreducible parts of a polymer chain that absorb and emit light. Fig.\ 5 illustrates the four LEGSs for a particular conformation of PPV with 40 moieties (i.e., 20 monomers).

Higher energy lying states are also localized, but are nodeful and generally spatially overlap a number of low-lying LEGSs. These states are named quasiextended exciton states (QEESs) and an example is  illustrated in Fig.\ 5.

\section{2.\ Fast C-C Bond Vibrations}\label{se:4.1}

The coupling of local normal modes (e.g., vinyl-unit bond stretches or phenyl-ring symmetric breathing modes) to a Frenkel exciton is conveniently described by the Frenkel-Holstein model\cite{Holstein59a,Barford14b},
\begin{eqnarray}\label{Eq:2}
  \hat{H}_{FH} &=& \hat{H}_{F}
 -A\hbar\omega_{\textrm{vib}}\sum_{n=1}^{N} \tilde{Q}_n \hat{N}_{n}
    + \frac{\hbar\omega_{\textrm{vib}}}{2}\sum_{n=1}^{N} \left( \tilde{Q}_n^2 +\tilde{P}_n^2 \right).
\end{eqnarray}
$\hat{H}_{F}$ is the Frenkel Hamiltonian, defined in Eq.\ (\ref{Eq:1}), while
  $\tilde{Q} = (K_{\textrm{vib}}/\hbar \omega_{\textrm{vib}})^{1/2}Q$
and
  $\tilde{P} =(\omega_{\textrm{vib}}/\hbar K_{\textrm{vib}})^{1/2}P$
are the dimensionless displacement and momentum of the normal mode.
The second term on the right-hand-side of Eq.\ (\ref{Eq:2}) indicates that the normal mode couples linearly to the local exciton density. $A$ is the dimensionless exciton-phonon coupling constant which introduces  the local Huang-Rhys factor $S = A^2/2$.
The final term is the sum of the elastic and kinetic energies of the harmonic oscillator, where
$\omega_{\textrm{vib}}$ and $K_{\textrm{vib}}$ are the angular frequency and force constant of the oscillator, respectively.

Exciton-nuclear dynamics is often modeled via the Ehrenfest approximation, which treats the nuclear coordinates as classical variables moving in a mean field determined by the exciton. However, (as discussed in the main text) the Ehrenfest approximation fails to correctly describe ultrafast dynamical processes.
A correct description of the coupled exciton-nuclear dynamics therefore requires a full quantum mechanical treatment of the system. This is achieved by introducing the harmonic oscillator raising and lowering operators, $\hat{b}_n^{\dagger}$ and $\hat{b}_n$, for the normal modes i.e.,
$ \tilde{Q}_n \rightarrow \hat{\tilde{Q}}_n = (\hat{b}_n^{\dagger} +  \hat{b}_n)/\sqrt{2}$
and
 $ \tilde{P}_n \rightarrow \hat{\tilde{P}}_n = i(\hat{b}_n^{\dagger} -  \hat{b}_n)\sqrt{2}$.
The time evolution of the quantum system can then conveniently be simulated via the TEBD method, as described in Section 5.

\subsection{Exciton-polarons}


Since the photoexcited system has a different electronic bond order than the ground state, an instantaneous force is established on the nuclei. This force creates an exciton-polaron, whose spatial size is quantified by the exciton-phonon correlation function\cite{Hoffmann02}
\begin{equation}\label{Eq:15}
C_n^{\textrm{ex-ph}}(t) \propto \sum_m \langle\hat{N}_{m} \hat{\tilde{Q}}_{m+n} \rangle.
\end{equation}
$C_n^{\textrm{ex-ph}}$ correlates the local phonon displacement, $Q$, with the instantaneous exciton density $n$ moieties away.
$C_n^{\textrm{ex-ph}}(t)$, illustrated
in Fig.\ 7 of ref\cite{Mannouch18},
shows that the exciton-polaron is established within 10 fs (i.e., within half the period of a C-C bond vibration) of photoexcitation. The temporal oscillations, determined by the C-C bond vibrations, are damped as energy is dissipated into the vibrational degrees of freedom, which acts as a heat bath for the exciton. The exciton-phonon spatial correlations decay exponentially, extending over ca.\ 10 moieties. This short range correlation occurs because the C-C bond can respond relatively quickly to the exciton's motion.

\section{3.\ Slow Bond Rotations}\label{se:4.3}

Assuming that bond rotations are not sterically hindered, their coupling to the excitons is conveniently modeled  by supplementing the Frenkel-Holstein model (i.e., Eq.\ (\ref{Eq:2})) by\cite{Barford18}
\begin{eqnarray}\label{Eq:8}
 \hat{H}_{\textrm{rot}} = - \sum_{n=1}^{N-1}
B(\theta_n^0)\times ({\phi}_{n+1} -  {\phi}_{n})\hat{T}_{n,n+1}
+ \frac{1}{2}  \sum_{n=1}^{N} \left( K_{\textrm{rot}}{\phi}_n^2 + {L}_n^2/I \right).
\end{eqnarray}
Here, ${\phi}$ is the  angular displacement of a  from its groundstate equilibrium value (see Fig.\ 4) and
${L}$ is the associated angular momentum of a  moiety around its bridging bonds. Eq.\ (\ref{Eq:8}) is derived in ref\cite{Barford18}, assuming that the angular displacements are small.

The first term on the right-hand-side of Eq.\ (\ref{Eq:8}) indicates that the change in the dihedral angle,
 $\Delta {\theta}_{n}= ( {\phi}_{n+1} -  {\phi}_{n})$,
couples linearly to the bond-order operator,
\begin{equation}
 \hat{T}_{n,n+1} = \sum_{n=1}^{N-1}( |n\rangle \langle n+1| +  |n+1\rangle \langle n|),
\end{equation}
where \begin{equation}\label{Eq:24}
B(\theta_n^0) = J_{SE}^0\sin 2 \theta_n^0
\end{equation}
is the  exciton-roton coupling constant and $\theta_n^0$ is the groundstate dihedral angle for the $n$th bridging bond.
The final term is the sum of the elastic and kinetic energies of the rotational harmonic oscillator.

The natural angular frequency of oscillation is $\omega_{\textrm{rot}} = (K_{\textrm{rot}}/I)^{1/2}$, where $K_{\textrm{rot}}$ is the elastic constant of the rotational oscillator and $I$ is the moment of inertia, respectively. $K_{\textrm{rot}}$ is larger for the bridging bond  in the excited state than the groundstate, because of the increase in bond order. Also notice that both the moment of inertia (and thus $\omega_{\textrm{rot}}$)  of a rotating moiety and its viscous damping from a solvent are strongly dependent on the side groups attached to it. As discussed in the next section, this observation has important implications for whether the motion is under or over damped and on its characterstic timescales.

\subsection{A single torsional oscillator}\label{se:4.3.1}

Before considering a chain of torsional oscillators, it is instructive to review the dynamics of a single, damped oscillator subject to both  restoring and  displacement forces. The equation of motion for the angular displacement is
\begin{equation}\label{}
  \frac{d^2\phi(t)}{dt^2} = -\omega_{\textrm{rot}}^2(\phi(t) - \phi_{\textrm{eq}}) -\gamma   \frac{d \phi(t)}{dt,}
\end{equation}
where $\phi_{\textrm{eq}} = \lambda/K_{rot}$ is proportional to the displacement force and $\lambda$ is defined in Eq.\ (\ref{Eq:14}).

In the \emph{underdamped regime}\cite{French71}, defined by $\gamma < 2\omega_{\textrm{rot}}$,
\begin{equation}\label{}
 \phi(t) = \phi_{\textrm{eq}}\left(1-\cos(\omega t) \exp(-\gamma t/2)\right),
\end{equation}
where $\omega = (\omega_{\textrm{rot}}^2 - \gamma^2/4)^{1/2}$. In this regime, the torsional angle undergoes damped oscillations with a period $T = 2\pi/\omega$ and a decay time $\tau = 2/\gamma$.

Conversely,  in the \emph{overdamped regime}\cite{French71}, defined by $\gamma > 2\omega_{\textrm{rot}}$,
\begin{eqnarray}\label{}\nonumber
 \phi(t) = \phi_{\textrm{eq}}\left(1- \frac{1}{4\beta}\left(\gamma_1\exp(-\gamma_2 t/2) - \gamma_2\exp(-\gamma_1 t/2)\right)  \right),
\end{eqnarray}
where $\gamma_1 = \gamma + 2\beta$,  $\gamma_2 = \gamma - 2\beta$ and $\beta = (\gamma^2/4 - \omega_{\textrm{rot}}^2)^{1/2}$.
Now, the torsional angle undergoes damped biexponential decay with the decay times $\tau_1 = 2/\gamma_1$ and $\tau_2 = 2/\gamma_2$. In the limit of strong damping, i.e., $\gamma \gg 2\omega_{\textrm{rot}}$, there is a fast relaxation time $\tau_1 = 1/\gamma = \tau/2$ and a slow relaxation time $\tau_2 = \gamma/\omega_{\textrm{rot}}^2 \gg \tau$. In this limit, as the slow relaxation dominates at long times, the torsional angle approaches equilibrium with an effective mono-exponential decay.

For a polymer without alkyl side groups, e.g., PPP and PPV, $\omega_{\textrm{rot}} \sim  \gamma \sim 10^{13}$ s$^{-1}$ and are thus  in the underdamped regime with sub-ps relaxation. However, polymers with side groups, e.g., P3HT, MEH-PPV and PFO, have a  rotational frequency up to ten times smaller and a larger damping rate, and are thus in the overdamped regime\cite{Wells08}.

\subsection{A chain of torsional oscillators}\label{se:4.3.2}

An exciton delocalized along a polymer chain in a chromophore couples to multiple rotational oscillators resulting in collective oscillator dynamics.
Eq.\ (\ref{Eq:24}) and Eq.\ (\ref{Eq:14}) indicate that torsional relaxation only occurs if the moieties are in a staggered arrangement in their groundstate, i.e., $\theta_n^0 = (-1)^n\theta^0$. In this case the torque acts to planarize the chain. Furthermore, since the torsional motion is slow, the self-trapped exciton-polaron  is `heavy' and in the under-damped regime becomes self-localized on a timescale of a single torsional period, i.e., $200 - 600$ fs. In this limit the relaxed staggered bond angle displacement mirrors the exciton density. Thus, the exciton is localized precisely as for a `classical' Landau polaron and is spread over $\sim 10$ moieties\cite{Barford18}.

The time-evolution of the staggered angular displacement, $\langle{\phi}_n\rangle\times(-1)^n$, is shown in Fig.\ 5 of ref\cite{Barford18} illustrating that these displacements reach their equilibrated values  after two torsional periods (i.e., $t > 400$ fs). The inset also displays the time-evolution of the exciton density, $\langle N_n \rangle$, showing exciton density localization after a single torsional period ($\sim 200$ fs).

We have described how exciton coupling to torsional modes causes a spatially varying planarization of the moieties that acts as a one-dimensional potential which self-localizes the exciton. The exciton `digs a hole for itself', forming an exciton-polaron\cite{Landau33}. Some researchers\cite{Westenhoff06}, however, argue that torsional relaxation causes an exciton to become more \emph{delocalized}.  A mechanism that can cause exciton delocalization occurs if the disorder-induced localization length is shorter than the intrinsic exciton-polaron size.
Then, in this case for freely rotating moieties, the stiffer elastic potential in the excited state causes a decrease both in the  variance of the dihedral angular distribution, $\sigma_{\theta}^2 = k_BT/K_{\textrm{rot}}$, and the mean dihedral angle, $\theta^0$.
This, in turn, means that the exciton band width, $|4J|$, increases and the diagonal disorder\cite{Barford14b}, $\sigma_J = J_{SE} \sigma_{\theta}\sin 2\theta^0$,
decreases. Hence, the disorder-induced localization,  $L_{\textrm{loc}} \sim (|J|/\sigma_J)^{2/3}$, increases.

\section{4.\ Open Quantum Systems}\label{se:4.2}

Dissipation of energy from an open quantum system arising from system-environment coupling is commonly described by a Lindblad master equation\cite{Breuer02}
\begin{equation}
\label{Eq:9}
\frac{\partial\hat{\rho}}{\partial {t}} = -\frac{i}{\hbar}\left[\hat{H},\hat{\rho}\right]-\frac{{\gamma}}{2}\sum_{n}\left(\hat{L}_{n}^{\dagger}\hat{L}_{n}\hat{\rho}+\hat{\rho}\hat{L}_{n}^{\dagger}\hat{L}_{n}-
2\hat{L}_{n}\hat{\rho}\hat{L}_{n}^{\dagger}\right),
\end{equation}
where $\hat{L}_{n}^{\dagger}$ and $\hat{L}_{n}$ are the Linblad operators, and  $\hat{\rho}$ is the system density operator.
In practice, a direct solution of the Lindblad master equation is usually prohibitively expensive, as the size of Liouville space scales as the square of the size of the associated Hilbert space.
Instead, Hilbert space scaling can be maintained by performing ensemble averages over quantum trajectories (evaluated via the TEBD method), where the action of the Linblad dissipator is modeled by quantum jumps.\cite{Daley14}

Here we assume that the C-C bond vibrations couple directly with the environment \cite{Mannouch18},
in which case  the Linblad operators are the associated raising and lowering operators (i.e., $\hat{L}_{n} \equiv \hat{b}_{n}$, introduced in Section 2). In addition,
\begin{equation}
\label{Eq:7}
\hat{H}_{FH}' =\hat{H}_{\text{FH}}+\frac{\hbar\gamma}{4}\sum_{n}\left(\hat{\tilde{Q}}_{n}\hat{\tilde{P}}_{n}+\hat{\tilde{P}}_{n}\hat{\tilde{Q}}_{n}\right).
\end{equation}

In practice, a direct solution of the Lindblad master equation is usually prohibitively expensive, as the size of Liouville space scales as the square of the size of the associated Hilbert space.
Instead, Hilbert space scaling can be maintained by performing ensemble averages over quantum trajectories (evaluated via the TEBD method), where the action of the Linblad dissipator is modeled by quantum jumps, as described in Section 5.\cite{Daley14}

\section{5.\ Numerical Techniques}

\subsection{Time Evolving Block Decimation (TEBD)}

Given the initial condition, $|\Psi(t=0) \rangle$, the state of the system is propagated in time via the evolution operator, i.e.,
\begin{equation}\label{eq:16}
|\Psi(t+\delta t) \rangle  = \exp(-i\hat{H}\delta t/\hbar)|\Psi(t) \rangle,
\end{equation}
where
\begin{equation}\label{Eq:11}
\hat{H} = \hat{H}_{FH}' +  \hat{H}_{\textrm{rot}}.
\end{equation}
Thus, since the C-C bond vibrations are quantized, this method avoids the failures of the Ehrenfest approximation. In particular, exciton-site decoherence and non-adiabatic state interconversion are correctly described.

We perform this procedure numerically using the Tensor Network Theory Library\cite{Al-assam17}, which implements the Time Evolving Block Decimation (TEBD) method\cite{Vidal03,Vidal04}.  Details of the TEBD method are described in Appendix B of ref\cite{Mannouch18}. Briefly, the TEBD method works by (1) representing $|\Psi \rangle$ as a matrix product state\cite{Schollwock11}, (2) expressing the evolution operator via a Trotter decomposition, and (3) compressing the action of the evolution operator on $|\Psi \rangle$ via a singular value decomposition (SVD). Time-dependent expectation values of relevant observables are determined via $|\Psi(t) \rangle $.

Importantly, this approach is `numerically exact' as long as the truncation parameter exceeds $2^S$, where $S$ is the entanglement entropy, defined by $S = -\sum_{\alpha} \omega_{\alpha} \textrm{ln}_2 \omega_{\alpha}$ and $\{\omega\}$ are the singular values obtained at the SVD. The TEBD method permits the electronic and nuclear degrees of freedom to be treated as quantum variables on an equal footing. It thus rectifies all of the failures of the Ehrenfest method  and it is not limited by the representation of the potential energy surface.

\subsection{TEBD and Ehrenfest dynamics}

Unlike C-C bond vibrations, being over 10 times slower torsional oscillations can be treated classically\cite{Barford18}. Furthermore, since we are now concerned with adiabatic relaxation on a single potential energy surface, we may employ the Ehrenfest approximation for these modes.
Thus, using Eq.\ (\ref{Eq:8}), the torque on each ring is
\begin{eqnarray}\label{Eq:25}
\nonumber
\Gamma_n &=& - \frac{\partial \langle \hat{H}_{rot} \rangle}{\partial {\phi}_n}\\
&=& -K_{\textrm{rot}}{\phi}_n +\lambda_n
\end{eqnarray}
where we define
\begin{equation}\label{Eq:14}
\lambda_n = B(\theta_{n-1}^0) \langle \hat{T}_{n-1,n} \rangle - B(\theta_{n}^0) \langle \hat{T}_{n,n+1} \rangle.
\end{equation}
Setting $\Gamma_n = 0$ gives the equilibrium angular displacements in the excited state as
${\phi}_n^{\textrm{eq}} = \lambda_n/K_{\textrm{rot}}$.
${\phi}_n$ is  subject to the Ehrenfest equations of motion,
\begin{equation}\label{Eq:18}
I \frac{d {\phi}_n}{d {t}} = {L}_n,
\end{equation}
and
\begin{eqnarray}\label{Eq:19}
\frac{d {L}_n}{d {t}}  =  \Gamma_n- \gamma L_n,
\end{eqnarray}
where the final term represents the damping of the rotational motion by the solvent. Eq.\ (\ref{Eq:18}) and (\ref{Eq:19}) are integrated using the velocity Verlet algorithm and the classical values of $\{\phi\}$ are fed into $\hat{H}_{\textrm{rot}}$. Our dynamics are in the underdamped regime.

\subsection{The quantum jump trajectory method}\label{Se:2.E.2}

Rather than solve the Linblad master equation, Eq.\ (\ref{Eq:9}), directly, we use the quantum jump trajectory method and average over quantum trajectories.  In this method the time evolution of a single trajectory is  determined by the effective Hamiltonian\cite{Daley14}
\begin{equation}
\label{eq:8}
\hat{H}_{\text{eff}}=\hat{H}-\frac{\textrm{i}{\hbar\gamma}}{2}\sum_{n}\hat{b}_{n}^{\dagger}\hat{b}_{n},
\end{equation}
where $\hat{H}$ is the system Hamiltonian given in Eq.~(\ref{Eq:11}), $\gamma$ is the dissipation parameter and $\hat{b}_{n}$ is the Lindblad operator for site $n$.
Evolving this single quantum trajectory, initially in state $|\Psi(t)\rangle$, under the effective Hamiltonian gives a trial state at time $t+\delta t$
\begin{equation}
\label{eq:trial}
|\Psi_{\text{trial}}(t+\delta t)\rangle = \hat{U}(t+\delta t, t)|\Psi(t)\rangle,
\end{equation}
where $\hat{U}(t+\delta t, t)$ is the  evolution operator
\begin{equation}\label{eq:U}
  \hat{U}(t+\delta t, t) = \exp(-\textrm{i}\hat{H}_\text{eff}\delta t/\hbar).
\end{equation}

As the effective Hamiltonian in Eq.~(\ref{eq:8}) is non-Hermitian, the time evolution in Eq.~(\ref{eq:trial}) does not conserve the quantum state's norm
\begin{equation}
\label{eq:norm}
\langle\Psi_{\text{trial}}(t+\delta t)|\Psi_{\text{trial}}(t+\delta t)\rangle = 1-\delta p,
\end{equation}
where $\delta p$ is the amount the norm has decayed over the time step $\delta t$. Using Eq.~(\ref{eq:norm}), the state of the single quantum trajectory at time $t+\delta t$ is determined probabilistically as:
\begin{itemize}
\item[(i)] With probability $1-\delta p$, the state at time $t+\delta t$ is
\begin{equation}
\label{eq:timestep1}
|\Psi(t+\delta t)\rangle = \frac{\hat{U}(t+\delta t, t)|\Psi(t)\rangle}{\sqrt{1-\delta p}}.
\end{equation}

\item[(ii)] With probability $\delta p$, the state at time $t+\delta t$ is
\begin{equation}
\label{eq:timestep2}
|\Psi(t+\delta t)\rangle =\frac{\hat{b}_{n}|\Psi(t)\rangle}{\sqrt{\langle\Psi(t)|\hat{b}_{n}^{\dagger}\hat{b}_{n}|\Psi(t)\rangle}},
\end{equation}
\end{itemize}
where $\hat{b}_{n}$ is the Lindblad operator for site~$n$. The site at which the Lindblad operator is applied in Eq.~(\ref{eq:timestep2}) is again determined probabilistically, where the probability associated with applying the Lindblad operator at site~$n$ is given by
\begin{equation}
P_{n} = \frac{\langle\Psi(t)|\hat{b}_{n}^{\dagger}\hat{b}_{n}|\Psi(t)\rangle}{\sum_{m}\langle\Psi(t)|\hat{b}_{m}^{\dagger}\hat{b}_{m}|\Psi(t)\rangle}.
\end{equation}

Averaging over a large number of these stochastically determined trajectories accurately reproduces the time-dependent observables associated with the Lindblad master equation.\cite{Daley14}

\subsection{Ensemble averaging}

We perform our calculations of the fluorescence anisotropy over an ensemble of 1,000 different polymer conformations achieved by assuming that the torsional angle, $\theta$, is a Gaussian random variable with a mean of $25^{\circ}$ and a standard deviation of $10^{\circ}$. We also assume that the probability of a cis-trans defect is 8\%.
Furthermore, to include a general approach to the decay of our initial high energy QEESs onto  LEGSs, our results include an averaging over 10 different QEESs. Finally, our method of treating open quantum systems, via the quantum jump method, requires an averaging over  different quantum trajectories per initial QEES. The accuracy of this method increases with the number of quantum trajectories. We have used 20 different trajectories per initial QEES.



\subsection{Parameters}

\begin{table}[h]
\begin{tabular}{|c | c |}
 \hline
 Parameter & Value \\ [0.5ex]
 \hline
$J_{\textrm{DA}}$ & -0.60 eV\\
 $J_{\textrm{SE}}^0$ & -1.68 eV\\
 $\epsilon_{\textrm{phenyl}}$ & 5.0 eV\\
 $\epsilon_{\textrm{vinylene}}$ & 4.0 eV\\
   $A$ & 4\\
$\theta^0$ & $25^{\circ}$\\
\hline
$T_{\textrm{vib}} = 2\pi/\omega_{\textrm{vib}}$ & 21 fs\\
$T_{\textrm{rot}} = 2\pi/\omega_{\textrm{rot}}$ & 210 fs\\
 $\gamma^{-1}$  & 100 fs \\
 \hline
 $\sigma_{\theta}$ & $10^{\circ}$\\
    $x_{\textrm{cis-trans}}$ & 0.8\\
 \hline
\end{tabular}
\caption{Parameters used in the modeling of exciton dynamics in PPV.}
\end{table}
\end{suppinfo}

\bibliography{review}

\end{document}